\documentstyle[12pt,epsfig]{article}
\newcommand{\CA}{{\cal A}}

\newcommand{\CC}{{\cal C}}
\newcommand{\CD}{{\cal D}}

\newcommand{\CF}{{\cal F}}

\newcommand{\CJ}{{\cal J}}

\newcommand{\CM}{{\cal M}}

\newcommand{\CP}{{\cal P}}

\newcommand{\CS}{{\cal S}}
\newcommand{\CT}{{\cal T}}

\newcommand{\CV}{{\cal V}}
\newcommand{\CW}{{\cal W}}

\newcommand{\ga}{\alpha}
\newcommand{\gb}{\beta}
\newcommand{\g}{\gamma}

\newcommand{\gl}{\lambda}
\newcommand{\gt}{\theta}

\newcommand{\gs}{\sigma}

\newcommand{\go}{\omega}
\newcommand{\gG}{\Gamma}
\newcommand{\gD}{\Delta}
\newcommand{\gL}{\Lambda}

\newcommand{\gS}{\Sigma}

\newcommand{\gO}{\Omega}
\newcommand{\gve}{\varepsilon}
\newcommand{\gvt}{\vartheta}

\newcommand{\gvp}{\varphi}

\begin{document}

\begin{center}
 {\bf RELATIVISTIC CALCULATION OF POLARIZATION OBSERVABLES
      IN $NN\to d\pi$ PROCESSES }

\vskip 5mm
          A. Yu. Illarionov$^\dag$ , G. I. Lykasov
\vskip 5mm

{\small
 {\it { Joint Institute for Nuclear Research,\\
 141980 Dubna, Moscow Region, Russia}}
\\
$\dag$ {\it E-mail: alexej@nusun.jinr.ru}
}
\end{center}

\vskip 5mm

\begin{center}
\begin{minipage}{150mm}
\centerline{\bf Abstract}
  A detailed analysis of processes of the type $NN\to d\pi$ is presented
  taking into account the exchange graphs of a nucleon and a pion.
  A large sensitivity of polarization observables to the off-mass shell
  effects of nucleons inside the deuteron is shown. Some of these
  polarization characteristics can change the sign by including these
  effects. The influence of the inclusion of a $P$-wave in the deuteron wave
  function is studied, too. The comparison of the calculation results of
  all the observables with the experimental data on the reaction
  $pp\to d\pi^+$ is presented.
\\[5mm]
\end{minipage}
\end{center}

\vskip 10mm

 \section{\bf Introduction}

 As known, pion production in $NN$ collisions, in particular the
 channel $NN\to d\pi$, has been investigated by many theorists and
 experimentalists over the last decades. An earlier study of this reaction
 \cite{begin} shows that the excitation of the $\Delta$-isobar is a crucial
 ingredient for explaining the observed energy dependence of the cross
 section. A lot of papers are based on multichannel Schr\"odinger equations
 with separable or local potentials \cite{schrodinger}. However,
 those studies were performed within the nonrelativistic approach. Early
 attempts to develop the relativistic approach were made in
 \cite{begin_rel}. Both the pole graph, i.e. one-nucleon
 exchange, and the rescattering graph presented below were calculated in
 those papers. As shown, this diagram should
 result in a dominant contribution to the cross section of the discussed
 process. By the calculation of this one, some approximations, in particular
 the factorization of nuclear matrix elements, neglect of recoil etc.,
 were introduced which lead to an uncertainty of the final results.
 A more careful relativistic study of the reaction $pp\to d\pi^+$ was made
 in \cite{locher}. The pole and rescattering graphs were
 shown to be insufficient to describe the experimental data; higher order
 rescattering contributions  should be taken into account. However, in this
 approach there was no successful description of all the polarization
 observables, especially the asymmetries $A_{y0}$, $iT_{11}$.
 Really, analyzing reactions of the type $NN\to d\pi$, there occurs a
 problem related to the off-mass shell effects of nucleons inside the
 deuteron. When the pion is absorbed by a two-nucleon pair or the deuteron,
 the pion energy is shared between two nucleons. So, for example, the
 relative momentum of the nucleon inside the deuteron increases at least
 by a value $\sim\sqrt{m\mu}=360 MeV$ if the rest pion is absorbed by
 the off-shell nucleon what corresponds to intra-deuteron distances of the
 order of $\sim 1/\sqrt{m\mu}\simeq 0.6 fm$. This means that the absorption
 process should be sensitive to the dynamics of the $\pi NN$ system at small
 distances. In this paper we concentrate mainly on the investigation of the
 role of these effects and the contribution of the $P$-wave of the deuteron
 wave function \cite{gross}. The sensitivity of all the polarization
 observables to these effects is studied, and it is shown that some
 polarization characteristics can change the sign by including the off-mass
 shell effects of nucleons inside the deuteron.

   The detailed covariant formalism of the construction of the relativistic
 invariant amplitude of the reaction $NN\to d\pi$ for this process are
 presented in chapter 2.
   We analyze in detail both the pole graph, one-nucleon exchange, and the
 triangle diagram, i.e. the pion rescattering graph, in sections 3.
 The inputs by this consideration, the covariant pseudoscalar $\pi NN$  and
 deuteron $d\to pn$ vertices, are discussed in detail.
   The discussions of the obtained results and the comparison with the
 experimental data are presented in chapter 4.
   The conclusion is presented in the last section 5.

 \section{\bf General Formalism}

\noindent
$\bullet~~${ \it Relativistic invariant expansion of the amplitude. } \\
   We start with the basic relativistic expansion of the reaction amplitude
 $NN\to d\pi$ using Itzykson-Zuber conventions \cite{zuber}.
 In the general case, the relativistic amplitude of the production of two
 particles of spins $1$ and $0$ by the interaction of two spin $1/2$
 particles has $6$ relativistic invariant amplitudes if all particles are
 on-mass shell and taking $P$-invariance into account.
 It can be written in the following form:
\begin{eqnarray}
%%%%%%%%%%%%%%%%%%%%%%%%%%%%% Fig.1 %%%%%%%%%%%%%%%%%%%%%%%%%%%%%%%%%%%%
\hspace{-1cm}
\begin{minipage}{5.2cm}
\begin{center}
\unitlength1cm
\begin{picture}(6,4)
\thicklines
\put(3,2){\circle{1.4}}
\put(0,0){\makebox(6,4){$\chi_\mu$}}
\thinlines
\multiput(1,1.5)(0,1){2}{\vector(1,0){0.8}}
\multiput(1.8,1.5)(0,1){2}{\line(1,0){0.7}}
\multiput(3.5,1.5)(0.3,0){4}{\line(1,0){0.15}}
\put(4.7,1.5){\vector(1,0){0.15}}
\thicklines
\put(3.5,2.51){\line(1,0){1.265}}
\put(3.55,2.45){\line(1,0){1.21}}
\put(4.6,2.385){\makebox(0.2,0.1)[lb]{$>$}}
% Notes
\put(0.5,2.8){\parbox[b]{2cm}{$\bar v_{\sigma_2}^{~r_2}(p_2)$}}
\put(0.5,1){\parbox[b]{2cm}{$u_{\sigma_1}^{~r_1}(p_1)$}}
\put(4.5,2.8){\parbox[b]{2cm}{$\xi^{~(\beta)}_\mu(d)$}}
\put(4.8,1.05){\parbox[b]{2cm}{$\varphi_\pi$}}
\end{picture}
\end{center}
\end{minipage}
%%%%%%%%%%%%%%%%%%%%%%%%%%%%%%%%%%%%%%%%%%%%%%%%%%%%%%%%%%%%%%%%%%%%%%%%
\hfill
 {\cal M}_{\gs_2,\gs_1}^{\beta}(s,t,u)=
 \left[ \bar v_{\gs_2}^{~r_2}(p_2)\chi^\mu_{r_2r_1}(s,t,u)
      u_{\gs_1}^{~r_1}(p_1) \right] \xi^{~(\beta)}_\mu(d)\gvp_\pi,
\label{IA1}
\end{eqnarray}
 where $u_{\gs_1}^{~r_1}(p_1)\equiv u_1$ and
 $\bar v_{\gs_2}^{~r_2}(p_2)\equiv\bar v_2$ are the
 spinor and anti-spinor of the initial nucleons with spin projections
 $\gs_1$ and $\gs_2$ and dirac indices $r_1$ and $r_2$, respectively;
 $\xi_\mu(d)$ is the deuteron polarization vector, $\gvp_\pi$ is the
 $\pi$-meson field; $s,t,u$ are the invariant Mandelstam variables:
\begin{equation}
 s = (p_1 + p_2)^2~~;~~t = (d - p_2)^2~~;~~u = (d - p_1)^2~~.
\label{Mvar}
\end{equation}

 This amplitude should be symmetrized over the initial nucleon states, and
 therefore it takes the form:
\begin{equation}
 \bar {\cal M}_{\gs_2,\gs_1}^{\beta}=\frac{1}{\sqrt{2}}\left[
 {\cal M}_{\gs_2,\gs_1}^{\beta}(s,t,u)+
        {(-1)}^\beta{\cal M}_{\gs_1,\gs_2}^{\beta}(s,u,t)\right]
 \label{IA2}
\end{equation}
 The second term in (\ref{IA2}), corresponding to the exchange of two
 nucleons, is equivalent to the exchange of the $t-$ and $u-$ variables.

 Using transformation properties of the wave functions, one can finds
 the transformation lows of the spinor amplitude $\chi^\mu_{r_2r_1}$.
 {\it The Lorenz-invariance } of matrix element under the Lorenz 
 transformation of all four-vectors $p'=\gL({\cal A})p$ leads to the
 following Lorenz transformation low of the spinor amplitude:
\begin{equation}
 \chi_{\ga\gb}^{\mu}(p_1,p_2;d,\pi) = {\cal S}_{\ga}^{~\ga'}({\cal A})
    \chi_{\ga'\gb'}^{\mu'}(p_1',p_2';d',\pi'){\cal S}_{\gb}^{~\gb'}
    ({\cal A}^{-1})\gL_{\mu'}^{~\mu}({\cal A}^{-1})~.
\label{INVAR1}
\end{equation}

With respect to discrete symmetries, we have from {\it P-invariance}
\begin{equation}
\chi^\mu(\vec p_1,\vec p_2;\vec d,\vec\pi)=
\eta_{\cal P}\g_0\chi^\mu(-\vec p_1,-\vec p_2;-\vec d,-\vec\pi)
\g_0g^{\mu\mu}~,
\label{INVAR2}
\end{equation}
where $\eta_{\cal P} = 
{\eta_1\eta_2 \over \eta_\pi\eta_d}(-1)^{s_d - s_1 - s_2}=(-1)$;
 $\eta_i, s_i-$ are internal parities and spins of particles.

{\it Time-reversal symmetry} leads to time-reversal spinor amplitude
$\chi_\mu^{\ga\gb}$
\begin{equation}
\chi^\mu_{\ga\gb}(\vec p_1,\vec p_2;\vec d,\vec\pi) = 
\eta_\CT\CT_{\ga\ga'}^{-1}
\chi_\mu^{\ga'\gb'}(-\vec p_2,-\vec p_1;-\vec\pi,-\vec d)\CT_{\gb'\gb}
g^{\mu\mu}~,
\label{INVAR3}
\end{equation}
where the time-reversal matrix $\CT=-i\g_5\CC.$

{\it The charge conjugation} describe the connection of the spinor amplitudes
$\chi$ for the process $NN\to d\pi$ and $\chi_\CC$ for the charge conjugation
process $\bar N\bar N\to \bar d\pi$:
\begin{equation}
\chi(\vec p_1,\vec p_2;\vec d,\vec\pi)=\eta_\CC\CC
\chi_\CC^t(\vec p_1,\vec p_2;\vec d,\vec\pi)\CC^{-1}~.
\label{INVAR4}
\end{equation}

 The amplitude $\chi_\mu$ for the process  $NN\to d\pi$
 can be expanded over six independent covariants, which can choice in such
 way that every of them satisfy the above properties. For this one we
 introduce the orthogonal system of four-vectors, one of them, $P$, is
 time-like, and other, $p, N$ and $L$ are space-like:
\begin{equation}
P = p_1 + p_2,~p = (p_1 - p_2)/2,~
N_\mu =  \gve_\mu (p' p P),~L_\mu = \gve_\mu (N p P)~.
\label{basis}
\end{equation}
Here the four-vector $p' = (d - \pi)/2.$ Then, one can get the whole system 
of orthogonal unit four-vectors $\{e_\mu^{(\gs)}\}_{\gs = 0}^3$.
Therefore, the spinor amplitude $\chi_\mu$ can be expanded over this unit
orthogonal system
\begin{eqnarray} 
  \fbox{\parbox[c][0.5\height]{11.67cm}{
$$
\chi_\mu = \chi_ie_\mu^{(i)} = \chi_1l_\mu + \chi_2n_\mu + \chi_3e_\mu~,~~
\chi_i = -\chi^\mu e_\mu^{(i)}~ = \g_5\left( a_i + b_i{\widehat l} \right)
$$
   } } 
\label{IA3}
\end{eqnarray}

 \section{Reaction Mechanism}

\noindent
$\bullet~~$ {\it One-nucleon exchange (ONE) and $\pi NN$-vertex.}    \\
   Within the framework of the one-nucleon exchange model, the amplitude
 $\chi_\mu$ can be written in a simple form:
\vspace{-0.75cm}
\begin{equation}\begin{array}{cc}
%%%%%%%%%%%%%%%%%%%%%%%%%%%%%% ONE %%%%%%%%%%%%%%%%%%%%%%%%%%%%%%%%%%%%%
\begin{minipage}{5cm}
\begin{center}
\unitlength1cm
\begin{picture}(5,3)
\multiput(1,1)(0,1.2){2}{\vector(1,0){1.1}}
\multiput(2.1,1)(0,1.2){2}{\line(1,0){0.9}}
\multiput(3,1)(0.3,0){6}{\line(1,0){0.15}}
\put(4.8,1){\vector(1,0){0.2}}
\put(3,1){\circle*{0.1}}
\put(3,2.23){\line(1,0){2}}
\put(3,2.17){\line(1,0){2}}
\put(4.8,2.1){\makebox(0.2,0.1)[lb]{$>$}}
\put(3,2.2){\circle*{0.25}}
\thinlines
\put(3,1){\vector(0,1){0.7}}
\put(3,1.7){\line(0,1){0.5}}
% Notes
\put(1,2.4){\parbox[b]{1cm}{$p_2$}}
\put(1,0.7){\parbox[b]{1cm}{$p_1$}}
\put(4.7,2.4){\parbox[b]{1cm}{$d$}}
\put(4.7,0.7){\parbox[b]{1cm}{$\pi$}}
\put(3.1,1.5){\parbox{1cm}{$n$}}
\put(2.8,2.5){\parbox[b]{1cm}{$\bar\Gamma_\mu$}}
\put(2.8,0.6){\parbox[b]{1cm}{$\Gamma_5$}}
\end{picture}
\end{center}
\end{minipage}
%%%%%%%%%%%%%%%%%%%%%%%%%%%%%%%%%%%%%%%%%%%%%%%%%%%%%%%%%%%%%%%%%%%%%%%%
&
 \chi_\mu = g^+ \bar\gG_\mu(d) \CS_\CF(n) \gG_5(n)~,
 \label{ONE1}
\end{array}\end{equation}
 where
 $\bar\gG_\mu(d)$ is the deuteron vertex $pn\to d$ with one off-mass
 shell nucleon, $\CS_\CF(n)=\left(\widehat n-m+i0\right)^{-1}$ is the
 fermion propagator and the value of the coupling constant is
 $g^+ = \sqrt2g~,~~g^2/4\pi = 14.7~$. The vertex $\bar\gG_\mu(d)$ can be
 related to the deuteron wave function ($\CD\CW\CF$) with the help of the
 following equation \cite{gunion}:
\begin{equation}
 \bar\Psi_\mu={\bar\gG_\mu\over n^2-m^2+i0}
             =\gvp_1(t)\g_\mu+\gvp_2(t){n_\mu\over m}+
 \left(\gvp_3(t)\g_\mu+\gvp_4(t){n_\mu\over m}\right){\widehat n-m\over m}~.
 \label{ONE2}
\end{equation}
 The form factors $\gvp_i(t)$ are related to two large components of
 the $\CD\CW\CF$ $u$ and $w$ (corresponding to the $^3\CS_1$ and $^3\CD_1$
 states) and to small components $v_t$ and $v_s$ (corresponding to the
 $^3\CP_1$ and $^1\CP_1$ states) as in \cite{gross}.

   In the theoretical description of processes at intermediate energies,
 the structure of hadrons is often described by multiplying the point-like
 operators by form factors. It is common practice to assume that these
 vertices, i.e. their operator structures and the associated form factors,
 are in all situations the same as for a free on-shell hadrons. In our case,
 however, the pion vertex can have a much richer structure: there can be
 more independent vertex operators and the form factors can depend on more
 than one scalar variable. The situation is similar to the construction of
 the off-shell electromagnetic vertex \cite{EMV}. The common treatment of 
 such off-shell effects is to presume them small and to ignore them by using 
 the free vertices. However, as much of the present effort in intermediate
 energy physics focuses on delicate effects, such as evidence of
 quark/gluon degrees of freedom or small components in the wave functions, it
 is mandatory to examine these issues in detail \cite{davidson}.

   The most general pion-nucleon vertex, where the incoming
 nucleon of mass $m$ has momentum $p_i^\mu$, the outgoing nucleon has
 momentum $p_f^\mu$ and the pion has momentum $\pi^\mu = p_f^\mu - p_i^\mu$,
 can be written as \cite{kazes}
\begin{equation}
 \Gamma_5\left(p_f, p_i\right) = \gamma_5 G_1 +
        \frac{\widehat p_f - m}{m} \gamma_5 G_2 +
        \gamma_5 \frac{\widehat p_i - m}{m} G_3 +
        \frac{\widehat p_f - m}{m} \gamma_5 \frac{\widehat p_i - m}{m} G_4~;
 \label{ONE4}
\end{equation}
 here $\{G_i(t; p_i^2, p_f^2)\}_{i=1}^4$ are some functions depending on the
 relativistic invariant transfer $t = (p_i - p_f)^2$ and particles masses
 $p_{i,f}^2$ or the so-called pion form factors. By sandwiching $\Gamma_5$
 between on-shell spinors one obtains
 $G_1(t,m^2,m^2) \bar u(p_f) \gamma_5 u(p_i)$.

   In our case, one nucleon is the off-shell only, and therefore we will
 consider the ``half-off-shell'' vertex with incoming nucleon on-shell. We
 obtain in that case two terms in eq.(\ref{ONE4}) instead of four because
 the third and the fourth ones are vanishing, taking into account the
 Dirac equation for a free fermion. Then, eq.(\ref{ONE4}) can be written in
 the form:
\begin{equation}
 \Gamma_5(t) = \gamma_5 \left( G_1(t) + G_2(t)\frac{\widehat n+m}{m}\right)~=
              \lambda G^{\CP\CS}(t) \gamma_5 + \left (1-\lambda \right)
                      G^{\CP\CV}(t) {\widehat\pi\over2m} \gamma_5~,
 \label{ONE5}
\end{equation}
 Note, according to the so-called equivalence theorem \cite{shweb} the
 sum of all Born graphs for elementary processes, for example the
 pion photoproduction on a nucleon \cite{scherer} and the other ones, is
 invariant under chiral transformation \cite{friar}. This means that starting
 with the Lagrangian appropriate to the pseudoscalar $(\CP\CV)$ coupling,
 one ends up in the Lagrangian appropriate to the pseudoscalar
 $(\CP\CS)$ coupling by performing a chiral transformation. This
 equivalence theorem is related to the processes for elementary
 particles. But in our case, for the reaction $NN\rightarrow d\pi$
 there is a bound state, a deuteron, and therefore reducing this process
 to the one where only elementary particles participate, we will have
 the diagrams of a higher order over the coupling constant
 than the Born graph. So, the equivalence theorem cannot be applied
 to our considered processes. Therefore, the vertex $\Gamma_5$
 in our case can be written in the form of eq.(\ref{ONE5})
 which is actually a linear combination of pseudoscalar and pseudovector
 coupling with the so-called mixing parameter $\gl$.

 The $dNN$ vertex has been studied by Buck and Gross \cite{gross} within the
 framework of the Gross equation of nucleon-nucleon scattering. They used
 a one boson exchange (OBE) model with $\pi, \rho, \go$ and $\gs$ exchange.
 In their study, they suggest that the form factors $G^{\CP\CS}$ and
 $G^{\CP\CV}$ have the same $t$ - dependence, in particular
 $G^{\CP\CS}(t) = G^{\CP\CV}(t) = h_N(t)$, and consider
 $\gl=0.0; 0.2; 0.4; 0.6; 0.8$ and $1.0$. In each case, the parameters of the
 OBE model were adjusted to reproduce the static properties of the deuteron.
 They found that the total probability of the small components of the
 $\CD\CW\CF$: $P_{small}=\int_0^\infty p^2dp\left[v_t^2(p)+v_s^2(p)\right]$,
 increases monotonically with growing $\gl$ from approximately $0.03\%$ for
 $\gl=0$ to approximately $1.5\%$ for $\gl=1$.

\vskip 5mm

 \noindent $\bullet~~$ {\it  Second-order graphs}\\
   Let us consider now the second order graph corresponding to the
 rescattering of the virtual $\pi$-meson by the initial nucleon.
 This mechanism of the $NN\to \pi d$ process has been analyzed by many
 authors, see, for example, \cite{begin_rel,locher}. Our procedure of the
 construction of the helicity amplitudes corresponding to the
 triangle graph is similar to the ones published by \cite{locher},
 and so we present it briefly. The most important result of this integration
 is the nucleon spectator contribution where the nucleon labelled $\eta$
 is on mass shell $(\eta^2 = m^2)$:
\begin{equation}
%%%%%%%%%%%%%%%%%%%%%%%%%%%%%% SECOND %%%%%%%%%%%%%%%%%%%%%%%%%%%%%%%%%%
\begin{minipage}{6cm}
\begin{center}
\unitlength1cm
\begin{picture}(5,3)
\put(0,2.2){\vector(1,0){0.8}}
\put(0.8,2.2){\line(1,0){0.7}}
\put(1.5,2.2){\vector(1,0){1.3}}
\put(2.8,2.2){\line(1,0){1.1}}
\put(1.5,2.2){\line(1,-1){0.36}}
\put(1.9,1.8){\vector(1,-1){0.36}}
\put(2.3,1.4){\line(1,-1){0.36}}
\put(2.7,1){\vector(1,1){0.7}}
\put(3.2,1.5){\line(1,1){0.7}}
\put(1.5,2.2){\circle*{0.1}}
\thicklines
\put(3.9,2.23){\line(1,0){1.5}}
\put(3.9,2.17){\line(1,0){1.5}}
\put(5.2,2.1){\makebox(0.2,0.1)[lb]{$>$}}
\put(3.9,2.2){\circle*{0.25}}
\thinlines
\put(0,1){\vector(1,0){0.8}}
\put(0.8,1){\line(1,0){1.9}}
\multiput(2.7,1)(0.45,0){5}{\line(1,0){0.35}}
\put(5,1){\vector(1,0){0.37}}
\put(2.7,1){\circle*{0.15}}
% Notes
\put(0.1,2.4){\parbox[b]{1cm}{$p_2$}}
\put(0.1,0.7){\parbox[b]{1cm}{$p_1$}}
\put(5,2.4){\parbox[b]{1cm}{$d$}}
\put(5,0.7){\parbox[b]{1cm}{$\pi$}}
\put(2.6,2.4){\parbox{1cm}{$\eta$}}
\put(3.5,1.5){\parbox{1cm}{$k$}}
\put(1.7,1.5){\parbox{1cm}{$q$}}
\put(1.3,2.45){\parbox[b]{1cm}{$\Gamma_5$}}
\put(3.7,2.45){\parbox[b]{1cm}{$\bar\Gamma_\mu$}}
\put(2.4,0.5){\parbox[b]{1cm}{$f_{\pi N}^{el}$}}
\end{picture}
\end{center}
\end{minipage}
%%%%%%%%%%%%%%%%%%%%%%%%%%%%%%%%%%%%%%%%%%%%%%%%%%%%%%%%%%%%%%%%%%%%%%%%
\hfill
 \chi_\mu^{sp}=
 {g^+\over (2\pi)^3}\int h_\pi(q^2){\CF_\mu\left(\vec\eta,
 \eta_0=\sqrt{\vec\eta^2+m^2}\right)\over q^2-\mu^2}{d^3\eta\over2\eta_0}
\label{SO1}
\end{equation}
 where $h_\pi(q^2)$ is the pion form factor corresponding to the off-mass
 shell $\pi$-meson in the intermediate state; a monopole form has been
 chosen $h_\pi(q^2) = (\gL^2 - \mu^2)/(\gL^2 - q^2)$
 as like as in \cite{machleidt}; here $\gL$ is the corresponding cut-off
 parameter. The general form of $\CF_\mu$ can be written as follows:
\begin{equation}
\CF_\mu =
  \Gamma_5 \CS_\CF^c(\eta) {\bar\Gamma}_\mu(d) \CS_\CF(k) f^{el}_{\pi N}~,
\label{SO2}
\end{equation}
 where $f^{el}_{\pi N}$ is the amplitude of $\pi N$ elastic scattering;
 it can be presented as expansion over two off-shell invariant amplitudes
 $f^{el}_{\pi N}=(A+B\widehat\pi)$ which depend on four momenta. We compute
 A and B from the on-shell $\pi N$ partial wave amplitudes
 $\CT_{l\pm}^{on}(s_{\pi N})$ under the assumption
\begin{equation}
  \CT_{l\pm}(s_{\pi N},t_{\pi N}, u_{\pi N}) \approx
  \CT_{l\pm}^{on}(s_{\pi N})~,
\label{SO3}
\end{equation}
 where $\CT_{l\pm}^{on}(s_{\pi N})$ are taken from the Karlsruhe-Helsinki
 phase shift analysis \cite{holer}. However, in the partial wave 
 decomposition of the invariant functions, full off-shell angular momentum 
 projectors are used for the lowest waves in the manner discussed for the 
 $NN\to NN\pi$ reaction in Ref.\cite{kroll}.

   The triple integral (\ref{SO1}) over azimuth $\gvp_\eta$, polar angle
 $\gvt_\eta$ and the magnitude of three-momentum $\eta$ must be done
 numerically for which we used a Gaussian quadrature. There are 6 triple
 integrals over a complicated complex integrand for each scattering angle.

    \section{Results and Discussions}

   In order to investigate the effect of small components of the $\CD\CW\CF$,
 we have calculated the differential cross section $d\sigma/d\Omega$,
 polarization characteristics $A_{ii}, A_{y0}$, etc. for $pp\to d\pi^+$
 as a function of scattering angle at proton kinetic energy $T_p=578 MeV$
 corresponding to pion kinetic one $T_\pi=147 MeV$ because at this
 energy the probability of $\gD$-isobar production by the two - step
 mechanism is rather sizeable. All the calculated quantities are in the
 Madison convention and compared with the experimental data \cite{arndt}
 and partial-wave analysis ($\CP\CW\CA$) by R. A. Arndt et al. \cite{said}
 (dotted curve). The cut-off parameter $\gL$ and the mixing one $\gl$
 corresponding to the $\pi NN$ vertex are chosen by the best fitting of the
 experimental cross section $d\gs/d\gO$ data. We have checked that the
 polarization curves change very little if we vary the cut-off parameter
 $\gL$.

   Note that the contribution of the triangle graph is very large at
 intermediate initial kinetic energies and much smaller at lower energies.
 It is caused by a large value of the cross section of elastic $\pi N$
 scattering because of a possible creation of the $\gD$-isobar at this
 energy. One can stress that the application of Locher's form $\CD\CW\CF$
 \cite{locher} does not allow one to reproduce the absolute value of the
 differential cross section (see Fig. 1.) over the whole region of
 scattering angle $\gvt$. But using the Gross approach for the $\CD\CW\CF$
 \cite{gross}, one can describe $d\gs/d\gO$ at $\gl=0.6-0.8$ rather well.

   The next interesting result which can be seen from Fig. (2-6) is a large
 sensitivity of all the polarization characteristics to the small components
 of the $\CD\CW\CF$. The asymmetry $A_{y0}$ (Fig. 2.) and the vector
 polarization $iT_{11}$ (Fig. 3.) calculated within the framework of Gross's
 approach particularly show this large sensitivity. These quantities are
 interference dominated and sensitive to the phases. The results for
 $iT_{11}$ have a wrong sign with Locher's form $\CD\CW\CF$ \cite{locher}. On
 closer inspection, we observe that the first term in eq.(\ref{Ob2}),
 $(\Phi^*_1-\Phi^*_3)\Phi_2$, is very big due to constructive interference
 $\Phi_1\approx -\Phi_3$. It is caused by the $N\triangle$ configuration in a
 relative $\CS$ wave having $pp$ spin zero ($^1\CD_2$ state). The $^1\CD_2$
 partial-wave dominates making $\Phi_{1,2,3}$ large, but the results are the
 same contribution to $\Phi^{\CJ=2}_1$ and $\Phi^{\CJ=2}_3$ (with opposite
 signs caused by the relevant Wigner d-function signature). Since the
 contribution of $\Phi_{4,5,6}$ is negligible, the sign problem for $iT_{11}$
 is therefore very sensitive to the $\Phi^{\CJ=0}_2$ (or $^1\CS_0$) partial
 wave. As $iT_{11}$ is very nearly proportional to $\Phi_2$, the phase of
 $\Phi_2$ determines the sign of $iT_{11}$.

   The right structure of the observables starts to appear gradually in the
 theoretical curves as one increases the mixing parameter $\gl$ in the
 Buck-Gross model, that is to say, as one increases the probability of the
 small components in the $\CD\CW\CF$. We have checked that this structure
 originates indeed from the small components $v_t$ and $v_s$ in
 eq.(\ref{ONE2}). If we make $v_t=v_s=0$ in the Buck-Gross model, then
 all curves become very similar to Locher's ones. Similarly, if we vary
 the $\pi NN$ vertex given by eq.(\ref{ONE5}) by considering $\gl$ between
 $0$ and $1$ but keep Locher's $\CD\CW\CF$, then the curves change
 very little again.

   The proton spin correlations $A_{ii}$ are presented in Fig. (4-6).
 Actually, the data on $A_{zz}$ (Fig. 4.) is the measure of the
 $\Phi_{4,5,6}$ magnitudes because the deviation of $A_{zz}$ from $-1$ is
 determined by these amplitudes (\ref{Ob1}). According to the partial wave
 decomposition, $\Phi_4$ and $\Phi_6$ are the amplitudes containing only
 triplet spin states in the $pp$ channel. One can conclude
 that the magnitudes of the spin-triplet amplitudes are somewhat small. As
 for $A_{yy}$ (Fig. 5.) and $A_{xx}$ (Fig. 6.), the terms proportional to
 $\Phi_1 + \Phi_3$ can be neglected because there is a phase relation
 $\Phi_1 \approx -\Phi_3$. Therefore, the deviation of $A_{yy}$ and $A_{xx}$
 from $-1$ is determined by $\Phi_{4,6}$ again, whereas $\Phi_5$ does not
 contribute to the numerator of $A_{yy}$.

   One can also see a large sensitivity of the observables $A_{ii}$ to the
 used form of $\CD\CW\CF$. The application of Gross's approach by the
 construction of $\CD\CW\CF$ \cite{gross} results in the shapes of these
 characteristics which are different from the corresponding ones obtained
 within the framework of Locher's approach \cite{locher}.

   Note, the energy dependence of all the observables within the framework of
 the suggested approach is the subject of our next investigation.

\section{Summary and Outlook}

   A relativistic model for the reaction $NN\to d\pi$ has been discussed in
 detail using two forms of the $\CD\CW\CF$ \cite{locher} and \cite{gross}. 
 One of them \cite{locher} was already used in the analysis of the 
 $pp\to d\pi$ process also taking into account the two-step mechanism with a
 virtual pion in the intermediate state.
   The difference between our approach and the model considered in
 \cite{locher} is the following. We have analyzed the sensitivity of all the
 observables to the form of $\pi NN-$current and the choice of the
 $\CD\CW\CF$ relativistic form. First of all, from the results presented in
 Fig. (1-6), one can see very large sensitivity of all the observables,
 especially of the polarization
 characteristics to the choice of the $\CD\CW\CF$ form. The inclusion of the
 $P$-wave contribution in the $\CD\CW\CF$ within the framework of Gross's
 approach \cite{gross} results in a better description of the experimental 
 data on the differential cross section and the polarization observables. 
 The next interesting result is related to the extraction of some new 
 information on the off-shell effects due to a virtual (off-shell) nucleon. 
 Comparing the observable with the experimental data (see Fig. (1-6)), one 
 can test the assumption, suggested by \cite{gross}, of a possible form of 
 the pion form factor and conclude that one cannot use the mixing parameter 
 $\gl=1$ as like as in \cite{locher}.

   One can stress that the one-nucleon exchange and the pion rescattering
 graphs have been studied only in this paper in order to investigate
 very important effects: off-mass shellness of nucleon and pion, and $P-$wave
 contribution to the $\CD\CW\CF$. The interactions in the initial $NN$ and
 final $d\pi$ states can be in principle contributed to the total amplitude
 of the considered reaction. However, it will be as a separate stage of this
 study because a more careful inclusion of elastic $NN$ and $d\pi$
 interactions at intermediate energies is needed.

 {\bf Acknowledgements.}

   We gratefully acknowledge very helpful discussions with
 E. A. Strokovsky, S. M. Dorkin, S. S. Semikh and F. Kleefeld.

    \section{\bf Appendix.}

 \noindent $\bullet~~${\it Helicity formalism.}   \\
   To calculate the observables, differential cross sections and
 polarization characteristics, it would be very helpful to construct the
 helicity amplitudes of the considered process $NN\to d\pi$. So, we use the
 helicity formalism for this reaction presented in Ref.\cite{soffer}.

   Let us introduce initial nucleon helicities $\mu_1,\mu_2$ and the final
 deuteron $\lambda$, and helicity amplitudes
 $\bar\CM^\lambda_{\mu_2,\mu_1}(W,\vartheta)$ depending on the initial
 energy $W$ in the $N-N$ c.m.s. and the scattering angle $\vartheta$
 analogous to \cite{locher}. This amplitude
 $\bar\CM^\lambda_{\mu_2,\mu_1}(W,\vartheta)$ corresponds to the transition
 of the $NN$ system from the state with helicities  $\mu_1,\mu_2=\pm 1/2$ to
 the state with $\lambda=\pm 1,0$. 

\noindent
 With respect to discrete symmetries, we have from {\it parity conservation}
 (\ref{INVAR2}):
\begin{equation}  
 \CM_{\mu_2\mu_1}^\lambda=
  \eta_P(-1)^{(\mu_2-\mu_1)-\lambda}\CM_{-\mu_2-\mu_1}^{-\lambda}=
  (-1)^{\mu_2+\mu_1+\lambda}\CM_{-\mu_2-\mu_1}^{-\lambda}~.
 \label{HA1}
\end{equation}
 {\it Time - reversal symmetry} (\ref{INVAR3}) leads to
\begin{equation}
 \CM_{\mu_2\mu_1}^\lambda=
  (-1)^{(\mu_2-\mu_1)-\lambda} \CM^{\mu_2\mu_1}_{\lambda}~.
 \label{HA2}
\end{equation}
 We use the abbreviations for helicity amplitudes as \cite{illar}.
 Using the expansion (\ref{IA3}), one can get the following form of the
 helicity amplitudes:
\begin{eqnarray}
 \Phi_{^1_3} = \bar\CM_{++}^\pm = \mp{1\over\sqrt2}{\gve\over m} \Big[
   a_1^s \cos\vartheta \pm ia_2^a - a_3^a \sin\vartheta\Big],~
   \Phi_2 = \bar\CM_{++}^0 =  {\gve\gve_d\over mM}\Big[
   a_1^s \sin\vartheta + a_3^a \cos\vartheta \Big],            \nonumber \\
 \Phi_{^4_6} = \bar\CM_{+-}^\pm = \mp{1\over\sqrt2}{p\over m} \Big[
   b_1^s \cos\vartheta \pm ib_2^s - b_3^a \sin\vartheta \Big],~
   \Phi_5 = \bar\CM_{+-}^0 =  {p\gve_d\over mM} \Big[
   b_1^s \sin\vartheta + b_3^a \cos\vartheta \Big],
~~\label{HA3}
\end{eqnarray}
 where $\chi_i^{\{_a^s\}}(s,t,u)$ are symmetric and antisymmetric
 combinations
 $\chi^{\{_a^s\} } = (\chi_i(\vartheta) \pm \chi_i(\pi - \vartheta))/\sqrt2$.
 All symmetry properties (\ref{HA1}) are satisfied by these amplitudes.

\vskip 5mm

  \noindent $\bullet~~${\it Observables. }   \\
   Using the helicity amplitudes (\ref{HA3}), one can calculate the all
 observables: differential cross section, asymmetry, deuteron tensor
 polarization and so on. Let us present now the expressions for the
 following observables in the c.m.s. using $\Phi_i$:
\begin{eqnarray}
 A_{y0}& = &4Im(\Phi_1\Phi^*_4+\Phi_2\Phi^*_5+\Phi_3\Phi^*_6)\gS^{-1},~~~
 A_{0y}(\gt)=A_{y0}(\pi-\gt),
\nonumber\\
 A_{xz}& = &-4Re(\Phi_1\Phi^*_4+\Phi_2\Phi^*_5+\Phi_3\Phi^*_6)\gS^{-1},~~~
 A_{zx}(\gt)=A_{xz}(\pi-\gt),
\nonumber \\
 A_{zz}& = &-1+4(|\Phi_4|^2+|\Phi_5|^2+|\Phi_6|^2)\gS^{-1},
\nonumber \\
 A_{yy}& = &-1+2(|\Phi_1+\Phi_3|^2+|\Phi_4+\Phi_6|^2)\gS^{-1},
\nonumber \\
 A_{xx}& = &A_{zz}+2(|\Phi_1+\Phi_3|^2-|\Phi_4+\Phi_6|^2)\gS^{-1}.
\label{Ob1}
\end{eqnarray}
 The expressions for the deuteron tensor polarization components are
the following:
\begin{eqnarray}
 iT_{11} & = &-\sqrt{6}Im\left [(\Phi^*_1-\Phi^*_3)\Phi_2+
              (\Phi^*_4-\Phi^*_6)\Phi_5 \right]\gS^{-1},
\nonumber \\
 T_{20} & = &\left[1-6(|\Phi_2|^2+|\Phi_5|^2)\gS^{-1}\right]/\sqrt{2},
\nonumber \\
 T_{21} & = &\sqrt{6}Re\left[(\Phi^*_1-\Phi^*_3)\Phi_2+
             (\Phi^*_4-\Phi^*_6)\Phi_5\right]\gS^{-1},
\nonumber \\
 T_{22} & = &2\sqrt{3}Re(\Phi^*_1\Phi_3+\Phi^*_4\Phi_6)\gS^{-1}=
             (1+3A_{yy}-\sqrt{2}T_{20})/(2\sqrt{3})~.
\label{Ob2}
\end{eqnarray}
 The variable $\Sigma$ is related to the differential cross section as
\begin{equation}
\Sigma = 2 \sum_1^6{\mid \Phi_i\mid}^2
       = 4\frac{p}{k}{(\frac{m}{4\pi\sqrt{s}})}^{-2} \frac{d\sigma}{d\Omega}
       = {1\over\sigma_0}\frac{d\sigma}{d\Omega}~,
\label{Ob3}
\end{equation}
where $p$ and $k$ are the momenta of initial proton and final deuteron
in the c.m.s.

%-----------------------------------------------------------------------

%---------------------------------------------------------------------------

\newpage

% ------------------ Figure 1------------------
\begin{figure}[t]
\centering
~\\[-1.3cm]
\epsfig{file=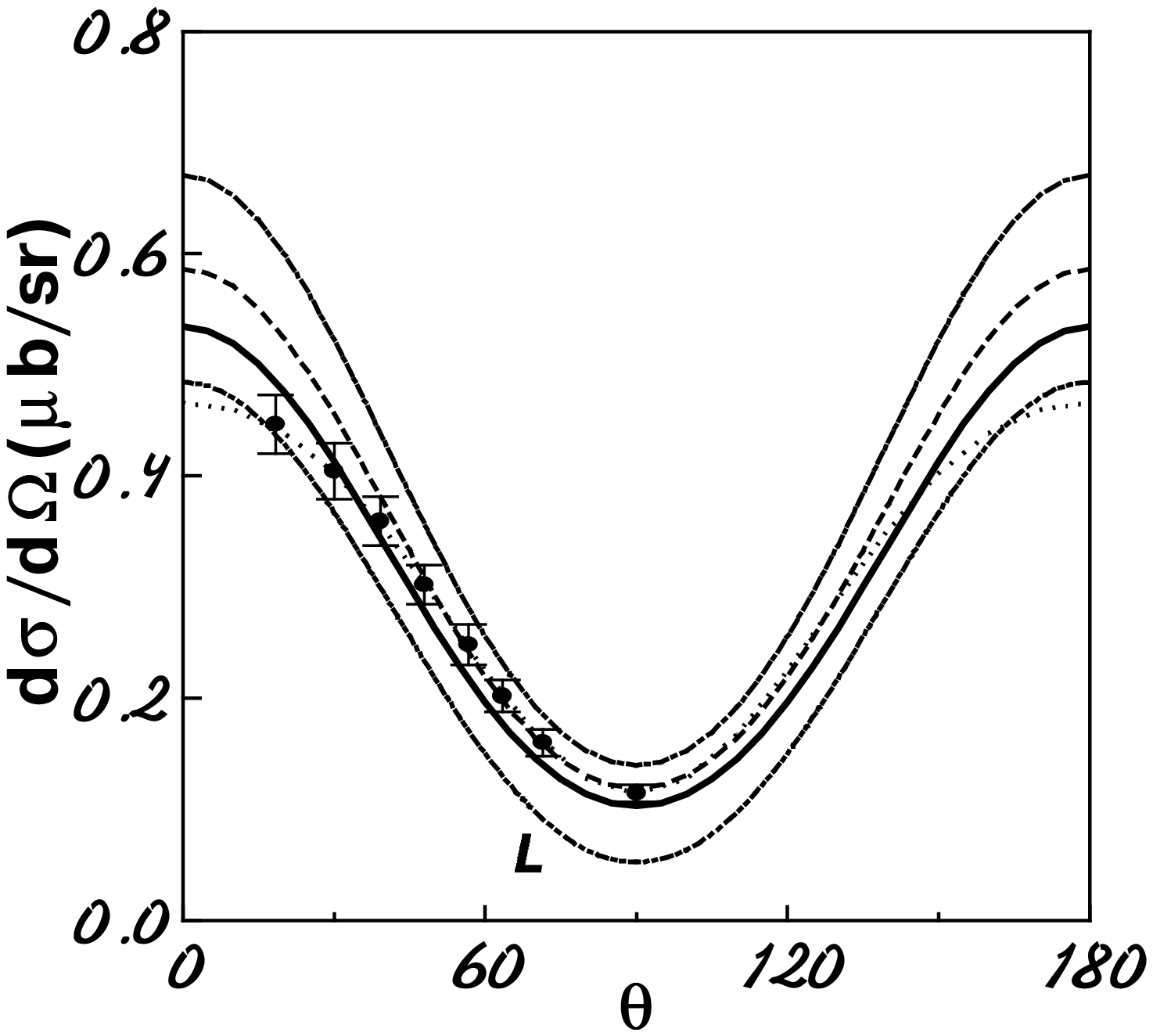,width=10.5cm}~~~~~~~
~\\[8.7cm]
\caption{
Differential cross section $d\gs/d\gO$
%,asymmetry $A_{y0}$ and vector polarization $iT_{11}$
for $pp\to d\pi^+$ as a function of scattering angle
in the c.m.s. at $T_p=578 MeV$ when the cut-off parameter $\gL$ and mixing
one $\gl$ varied simultaneously both in the deuteron wave function and in
the $\pi NN$ vertex. The dashed ($\gl=0.6; \gL=1$), solid ($\gl=0.8;
\gL=0.6$) and dot-dashed ($\gl=1;\gL=0.6$) lines correspond to the Gross
$\CW\CF\CD$ \protect\cite{gross}. The dot-dot-dashed line corresponds to the
results with Locher's $\CW\CF\CD$ \protect\cite{locher} ($\gl=1; \gL=1$).
The dots represent the partial-wave analysis by R. A. Arndt et al.
\protect\cite{arndt}. The data are from \protect\cite{locher,arndt}.
All spin observables are in the Madison convention.
}
%\label{cross}
\end{figure}
% ---------------------------------------------
%\vfill
% ------------------ Figure 2------------------
\begin{figure}[b]
\centering
%~\\[-2cm]
\epsfig{file=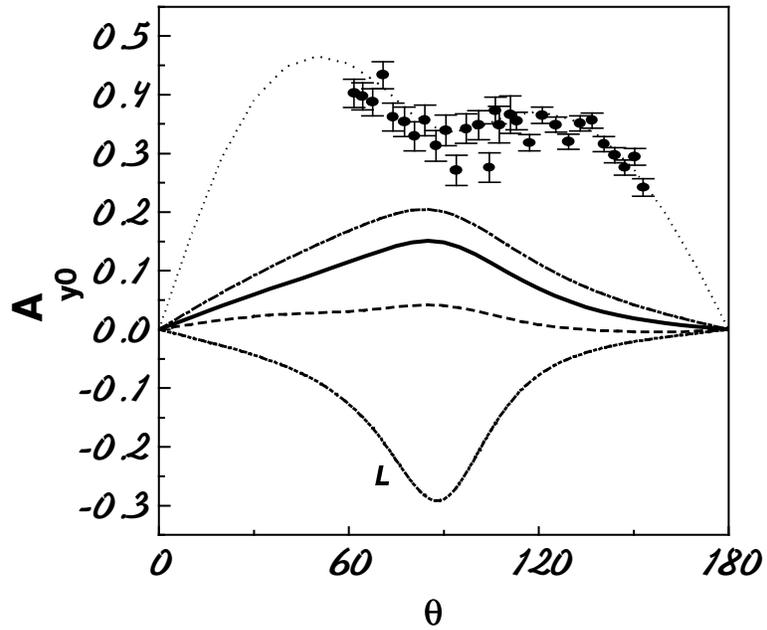,width=10.5cm}~~~~~~~
~\\[10.3cm]
\caption{Assymetry $A_{y0}$. Notation as in Fig. 1.}
%\label{ay0}
\end{figure}
% ---------------------------------------------
% ------------------ Figure 3------------------
\begin{figure}[t]
\centering
~\\[-1cm]
\epsfig{file=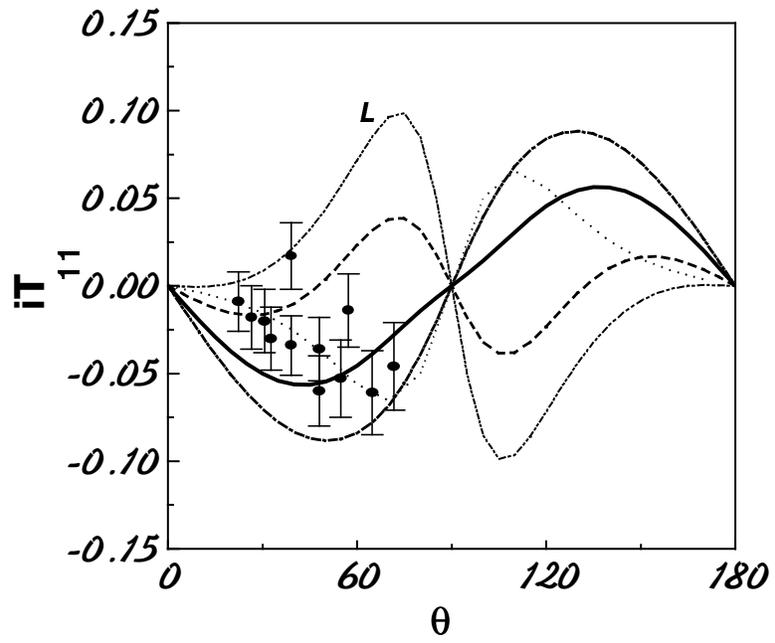,width=10.5cm}~~~~~~~
~\\[10.6cm]
\caption{Vector polarization $iT_{11}$. Notation as in Fig. 1.}
%\label{it11}
\end{figure}
% ---------------------------------------------
% ------------------ Figure 4------------------
\begin{figure}[b]
\centering
~\\[-1cm]
\epsfig{file=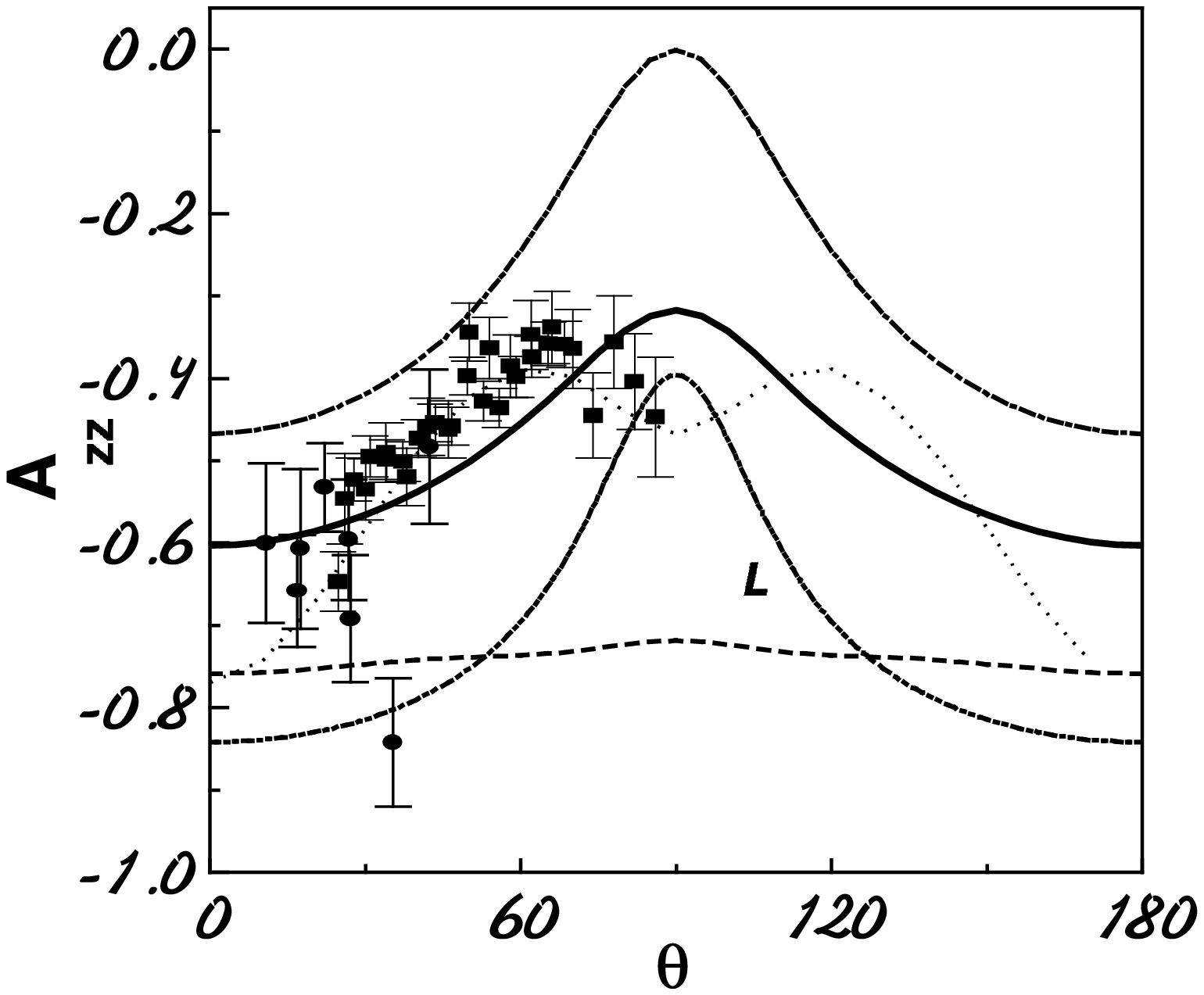,width=10.5cm}~~~~~~~
~\\[10.6cm]
\caption{Spin correlation $A_{zz}$. Notation as in  Fig. 1.}
\label{axx}
\end{figure}
% ---------------------------------------------

\newpage

% ------------------ Figure 5------------------
\begin{figure}[t]
\centering
~\\[-1cm]
\epsfig{file=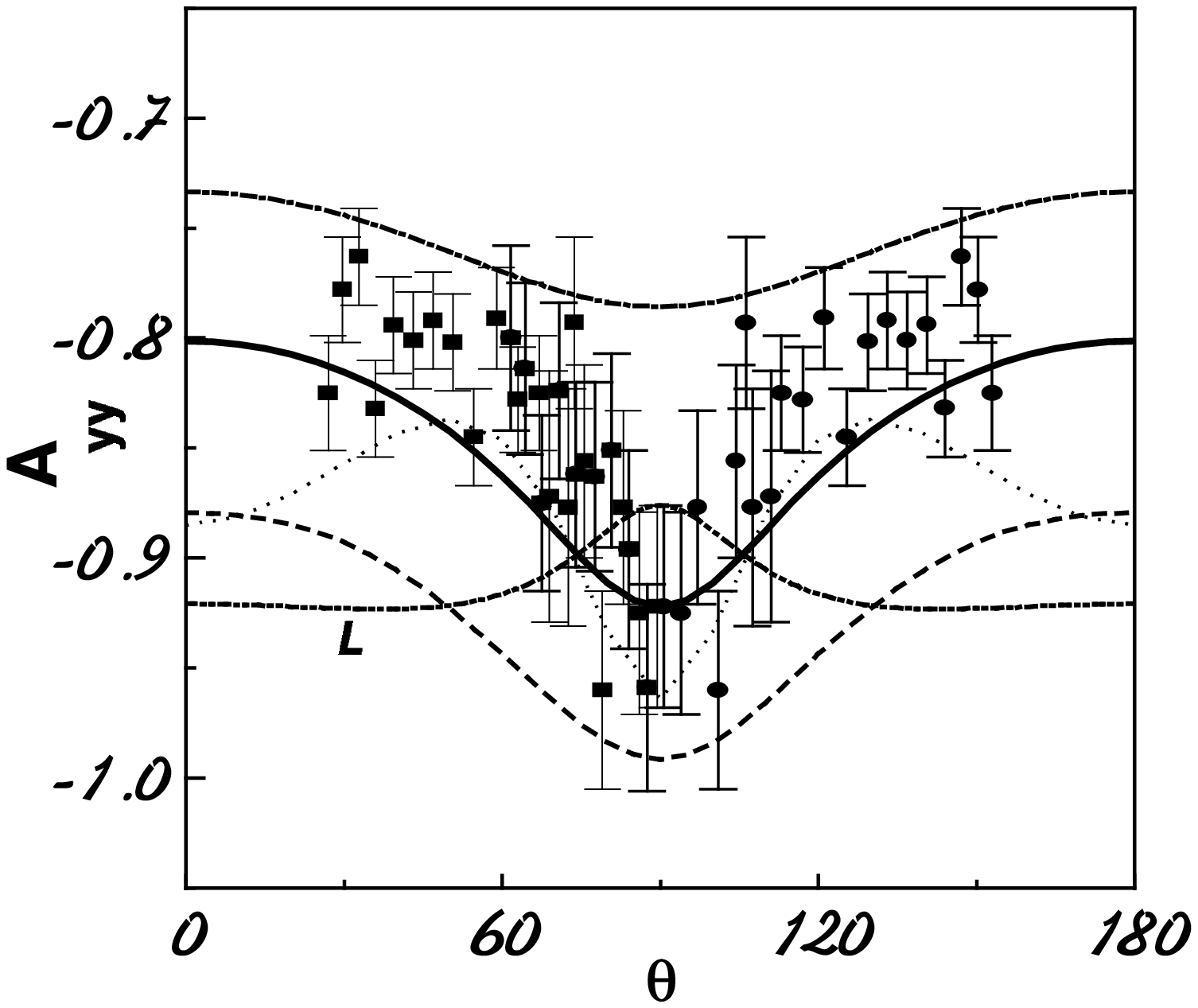,width=10.5cm}~~~~~~~
~\\[10.6cm]
\caption{Spin correlation $A_{yy}$. Notation as in Fig. 1.}
%\label{ayy}
\end{figure}
% ---------------------------------------------
% ------------------ Figure 6------------------
\begin{figure}[b]
\centering
~\\[-1cm]
\epsfig{file=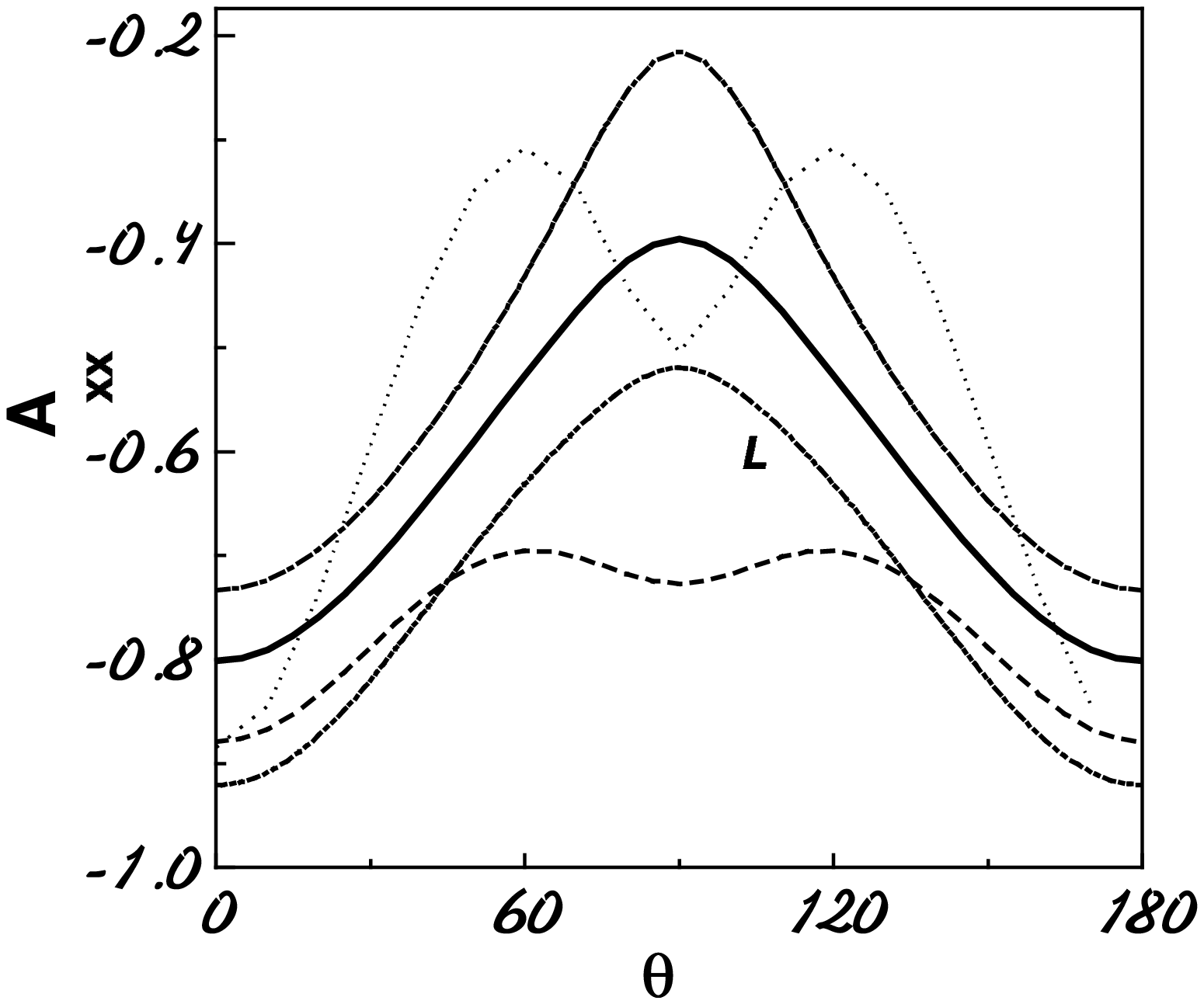,width=10.5cm}~~~~~~~
~\\[10.6cm]
\caption{Spin correlation $A_{xx}$. Notation as in Fig. 1.}
%\label{azz}
\end{figure}
% ---------------------------------------------

\end{document}